\documentclass[a4paper,10pt]{article}
\usepackage[a4paper]{geometry}

\usepackage{amsmath}
\usepackage{amssymb}
\usepackage{xcolor}
\usepackage{comment}
\usepackage[T1]{fontenc}
\usepackage[utf8]{inputenc}
\usepackage[english]{babel}
\usepackage{cite}
\usepackage{soul}

\usepackage{graphicx}
\usepackage[colorlinks=true,citecolor=blue,urlcolor=blue,linkcolor=blue]{hyperref}
\usepackage{amsthm}
\newtheorem{conjecture}{Conjecture}

\newtheorem{proposition}{Proposition}

\definecolor{lightblue}{rgb}{0.8,0.8,1}
\definecolor{OliveGreen}{cmyk}{0.44,0,0.95,0.20}

\title{Two-Unitary Complex Hadamard Matrices of Order $36$}
\author{Wojciech Bruzda$^{1}$\footnote{w.bruzda@uj.edu.pl}, Karol Życzkowski$^{1,2}$}
\date{{\small $^1$Centrum Fizyki Teoretycznej, Polska Akademia Nauk, Al. Lotników 32/46, 02-668 Warszawa\\
$^2$Instytut Fizyki Teoretycznej, Uniwersytet Jagielloński, ul. St. {\L}ojasiewicza 11, 30-348 Krak\'{o}w
\medskip
\\
{\small May 8, 2024}}
}

\begin{document}
\maketitle

\begin{abstract}
A family of two-unitary complex Hadamard matrices (CHM) stemming from a particular matrix,
of size $36$ is constructed.
Every matrix in this orbit remains unitary after operations of
partial transpose and reshuffling which makes it
a distinguished subset of CHM.
It provides a novel solution to the quantum version of the Euler
problem, in which each field of the Graeco-Latin square of size six contains a symmetric
superposition of all $36$ officers with phases being multiples of
sixth root of unity. This simplifies previously known solutions
as all amplitudes of the superposition are equal and the set of phases 
consists of $6$ elements only.  Multidimensional  parameterization
allows for more flexibility in a potential experimental realization.
\end{abstract}

\section{Introduction}

A unitary matrix $U$ of order $d^2$ is called two-unitary ($2$-unitary),
if the partially transposed matrix $U^{\rm\Gamma}$ and  the reshuffled matrix
$U^{\rm R}$ are also unitary.
Operations of reshuffling and partial transpose of any square-size matrix $X\in\mathbb{C}^{d^2\times d^2}$, addressed by a four-index $_{jk;lm}$, are defined as
\begin{equation}
X_{jk;lm}^{\rm R}=X_{jl;km}\quad\text{and}\quad 
X_{jk;lm}^{\rm\Gamma}=X_{jm;lk},
\end{equation}
where $X_{jk;lm}=\langle jk|X|lm\rangle$ is a representation of $X$ in
a local basis $\{|jklm\rangle\}$ of four copies of the Hilbert space; $\bigotimes_{j=1}^4\mathbb{C}^d=A\otimes B\otimes C\otimes D$~\cite{GALRZ15}. We use Dirac notation, where $|ab\rangle=|a\rangle\otimes|b\rangle$.
Two-unitary matrices play an significant role in the 
theory of quantum information.
They are related to quantum orthogonal Latin
squares~\cite{MV19}, perfect tensors~\cite{PYHP15} and absolutely maximally entangled states~\cite{GALRZ15}.
In the paper, we put the main focus on the last class of objects. In fact, any $2$-unitary matrix $U\in\mathbb{U}(d^2)$ corresponds to an absolutely maximally entangled state of four qudits (quantum  systems with $d$ degrees of freedom), written $|\psi\rangle\in{\rm AME}(4,d)$. Formally, state $|\psi\rangle$ is defined by the formula ${\rm Tr}_{Q'}|\psi\rangle\langle\psi|\propto\mathbb{I}$, where $Q$ denotes any balanced bipartition of four parties; $Q\cup Q'=\{A,B,C,D\}$, $Q\cap Q'=\emptyset$. Such states contain maximal entanglement with respect to any bipartition and provide important resource for many practical applications.

It is known \cite{HS00}
that $2$-unitary matrices do not exist for $d=2$. For larger dimensions, $d\ne 6$, $2$-unitary permutation matrices are implied by orthogonal Latin squares (OLS) of size $d$. Recent result 
\cite{RBBRLZ22,ZBRBRL23}
concerning the
particular case $d=6$ 
shows that such objects exist
for any $d>2$.
The original solution of the problem 
for a local dimension $d=6$ was shown not to be unique \cite{RRKL23},
which encourages us to search for further simplifications.

Throughout the paper, if not stated otherwise, the global dimension $N$ is always a square of the local dimension $d$ with $d^2 = N$, and we mostly focus on $d = 6$. By $\mathbb{N}_k=\{k, k+1, k+2, ...\}$ we denote the set of natural numbers starting with $k$, where $k\in\{0, 1, 2, ...\}$.
The three main sets of matrices we use are:
1) the set of $2$-unitary matrices, $\mathbb{U}^{\rm 2}(d^2)$,
2) complex Hadamard matrices $\mathbb{H}(d^2) = \{H\in\mathbb{U}(d^2) : \frac{1}{d^2}HH^{\dagger}=\mathbb{I}_{d^2}, |H_{jk}|=1\}$,
and 3) its proper subclass of Butson-type CHM~\cite{Bu62,Bu63},
\begin{equation}
{\rm B}\mathbb{H}(d^2, q)=\Big\{H\in\mathbb{H}(36) : H_{jk}=\exp\frac{i2\pi m_{jk}}{q}\Big\},
\end{equation}
for some $q\geqslant 2$ and $m_{jk}\in\mathbb{N}_0$.
Finally, let us define $\mathbb{H}^{\rm 2}(d^2) = \mathbb{H}(d^2)\cap\mathbb{U}^{\rm 2}(d^2)$.

The intersection of the sets of $2$-unitary matrices and CHM is non-empty.
For instance,  for $d=3$ one can notice~\cite{BRZ23} that
a suitably permuted tensor product 
of two Fourier matrices $F_3$
of order three is $2$-unitary,
\begin{equation}
(F_3\otimes F_3)P_9\in \mathbb{H}^{\rm 2}(9).
\end{equation}
Here $P_9$ denotes a permutation matrix of order nine, which determines the AME$(4,3)$ state of four subsystems with $3$ levels each~\cite{HC13}. An analogous construction can be used to construct $2$-unitary matrix of dimension $d^2$ for any dimension $d$ for which OLS($d$) exist. Such combinatorial designs are known for any $d>3$ apart from the $6$-dimensional Euler case~\cite{BSP60,Stin84}.

In this paper we answer the question about existence of a $2$-unitary complex Hadamard matrix of order $36$ affirmatively. This solution can be considered interesting from the quantum physics point of view, as it leads to a four-party state with a large coherence with respect to a generic locally equivalent basis. As an additional benefit, we provide not only a single representative of such a matrix, $\mathcal{H}$, but also a multidimensional affine family stemming from $\mathcal{H}$. This internal parameterization is different from the one considered in Ref.~\cite{RRKL23} as it preserves both properties of being $2$-unitary and Hadamard at the same time, and might serve as an independent tool in classification complex Hadamard matrices of square size and $2$-unitary matrices.

The paper is structured as follows: We start with recalling the algorithm that produces numerically $2$-unitary matrices of arbitrary dimension.
In Section~\ref{sec:2uCHM} we present analytical form of a $2$-unitary
complex Hadamard matrix $\mathcal{H}$ and its properties.
The full form of a family stemming from $\mathcal{H}$ is relegated to Appendix~\ref{app:full-form}, due to its algebraically overcomplex form.
Finally, in Section~\ref{sec:bu}, we compare our result
with recent work on biunimodular vectors which can be also used to construct
representatives of the set $\mathbb{H}^{\rm 2}(36)$.
Conclusions and future prospects are envisioned in the last Section~\ref{sec:summary}.

\section{\texorpdfstring{Sinkhorn algorithm}{}}

Let us briefly recall that originally a $2$-unitary representation $U\in\mathbb{U}(36)$ of the ``golden'' AME$(4,6)$ state was obtained numerically in Ref.~\cite{RBBRLZ22} using the iterative procedure $X_{t+1} = \mathcal{M}(X_t)$ for $t\in\mathbb{N}_0$ with
\begin{equation}
    \mathcal{M}(X) := \Pi\left(X^{\rm R\Gamma}\right),
\end{equation}
where $\Pi$ denotes polar decomposition projection onto the manifold of unitary matrices of order $N=36$.
This hardly converging procedure becomes to work nicely when
$\mathcal{M}$ is supplied with very particular seeds $X_0$ -- initial matrix points.
A collection of potential seeds consists of slightly perturbed permutation matrices. For example $X_0=P \exp(i\,\eta\,G)$, where $G$ is a random matrix whose entries are drawn according to the normal distribution. Matrix $P$ should be close\footnote{By the closeness one informally understands 
the number of transpositions (two-element cycles) that is required
to obtain a given permutation from another.}
to the permutation matrix $P_*$ defined as the best classical approximation of two orthogonal Latin squares~\cite{CGSS05}. Consequently, adding a non-zero ``noise'' controlled by small values of $\eta\in(0,1)$, guarantees the uniqueness of polar decomposition because the tiny perturbations provide an input for $\Pi$ that remains a full-rank matrix after operations of reshuffling and partial transpose.
Then, after several dozens of steps, e.g. $t\in\mathbb{N}_{256}$, one obtains
a fixed point of this map, which is the numerical approximation of a $2$-unitary matrix within the limits of the machine precision.
Final analytic shape of $U$ was a result of the tedious work
of searching for local unitary operations $V_j\in\mathbb{U}(6)$
such that $(V_1\otimes V_2)U(V_3\otimes V_4)$ took sparse enough form and allowed for its elements to be expressed as roots of unity located at the particular concentric circles around the origin.

Although the details and the general behaviour of this algorithm
are still not understood completely, it is possible to amend this procedure
to obtain even more interesting results.
One possible modification consists of adding additional step
which might be considered as purging the matrix elements from the noise.
To this end we define a
chopping procedure by means of the map
$c_{\varepsilon} : \mathbb{C}^{N\times N}\ni X\mapsto c_{\varepsilon}(X)\in\mathbb{C}^{N\times N}$, where
$c_{\varepsilon}(X_{jk})=0$ if $|X_{jk}|\leqslant\varepsilon$,
otherwise matrix elements stay intact.
This means, that near-zero entries are set to zero, which in general obviously
breaks unitarity, however, starting with a relatively small value of $\varepsilon\approx 0$ and gradually increasing it to $\varepsilon\nearrow 1$, one can smoothly steer the form of the final matrix. This is because above some threshold $\varepsilon>\varepsilon^*$, the operation $c_{\varepsilon}$ stops
affecting values of the matrix and, additionally, in some cases, the map $\mathcal{M}$
does not disturb its zero values either.
Formal explanation of these facts is currently beyond the scope of this report.
Here, we take the numerical behavior as a strong although obscured evidence
of a yet-to-be-discovered feature of this algorithm and,
define a new iterative procedure as
\begin{equation}
X_{t+1}= c_{\varepsilon}\left(\mathcal{M}\left(X_t\right)\right)=\mathcal{M}_{\varepsilon}(X_t)
\end{equation}
for $t\in\mathbb{N}_0$ and $\varepsilon\in(0,1]$.
We shall use a short notation $Y=\mathcal{M}_{\varepsilon}(X)$ to denote output $Y$ for a seed $X$.
This additional modification can significantly change possible outputs of the original map $\mathcal{M}$ and result in new analytical representatives of AME$(4,6)$ states. One must remember
that not every seed provides a solution, and still a kind of fine-tuning must be performed to make this recipe, based on the original algorithm of Sinkhorn~\cite{Si64, SK67}, converge quickly.

From now on, we fix $d=6$ and $N=d^2=36$. Again, the best seeds for $\mathcal{M}_{\varepsilon}$ are seemingly those which are close to the permutation matrix $P_*$~\cite{CGSS05}.
Provided that there exists $Y=\mathcal{M}_{\varepsilon}(X)$, for some seed $X$ (meaning that the procedure is convergent), the closer to $P_*$ the larger probability to obtain a more sparser form of $Y$, which is usually permutationally equivalent
to the block diagonal matrix with three blocks, each of size twelve.
In other words, there exist two permutation matrices $P_L$ and $P_R$
such that $P_LYP_R = \mathbb{B}_1\oplus\mathbb{B}_2\oplus\mathbb{B}_3$,
where $\mathbb{B}_j$ is one such block of size $12$.

Having a collection of outputs, $\mathbb{Y}=\{Y : Y=\mathcal{M}_{\varepsilon}(X) \ \text{with} \ X \in \mathbb{C}^{36\times 36}\}$, we ask whether some of them might be representatives of a CHM or become so when subjected to local unitary rotations. Paradoxically, this time we do not intend to make further simplifications but, in some sense, we are going to slightly ``complicate'' the matrix form, turning all entries (including zeros) into unimodular complex numbers. The objective function reads
\begin{equation}
    \mathcal{Z}(Y) = \min_{V_j\in\mathbb{U}(6)}\Bigg\{\sqrt{\sum_{j=1}^{36}\sum_{k=1}^{36}\Big(|(V_1\otimes V_2)Y_{jk}| - 1\Big)^2}\Bigg\},\label{Z-function}
\end{equation}
where optimization is performed over two local unitary matrices $V_1,V_2\in\mathbb{U}(6)$.
Mildly interesting fact is that we do not need
to introduce all four local unitaries, setting $V_3=V_4=\mathbb{I}$. 
Matrices $V_1$ and $V_2$ are initially drawn at random and during the optimization process at each step they are being converted to unitary matrices via polar decomposition; $V_j\mapsto \Pi(V_j+\Delta_j)$ for $j=1,2$ and perturbation $\Delta_j$ proportional to $\mathcal{Z}(Y)$.
Many numerical investigations have not exposed the necessity of using
additional pair, so only two of them, on either side, form a good enough structure.

\begin{figure}[!ht]
  \center
  \includegraphics[width=5.2in]{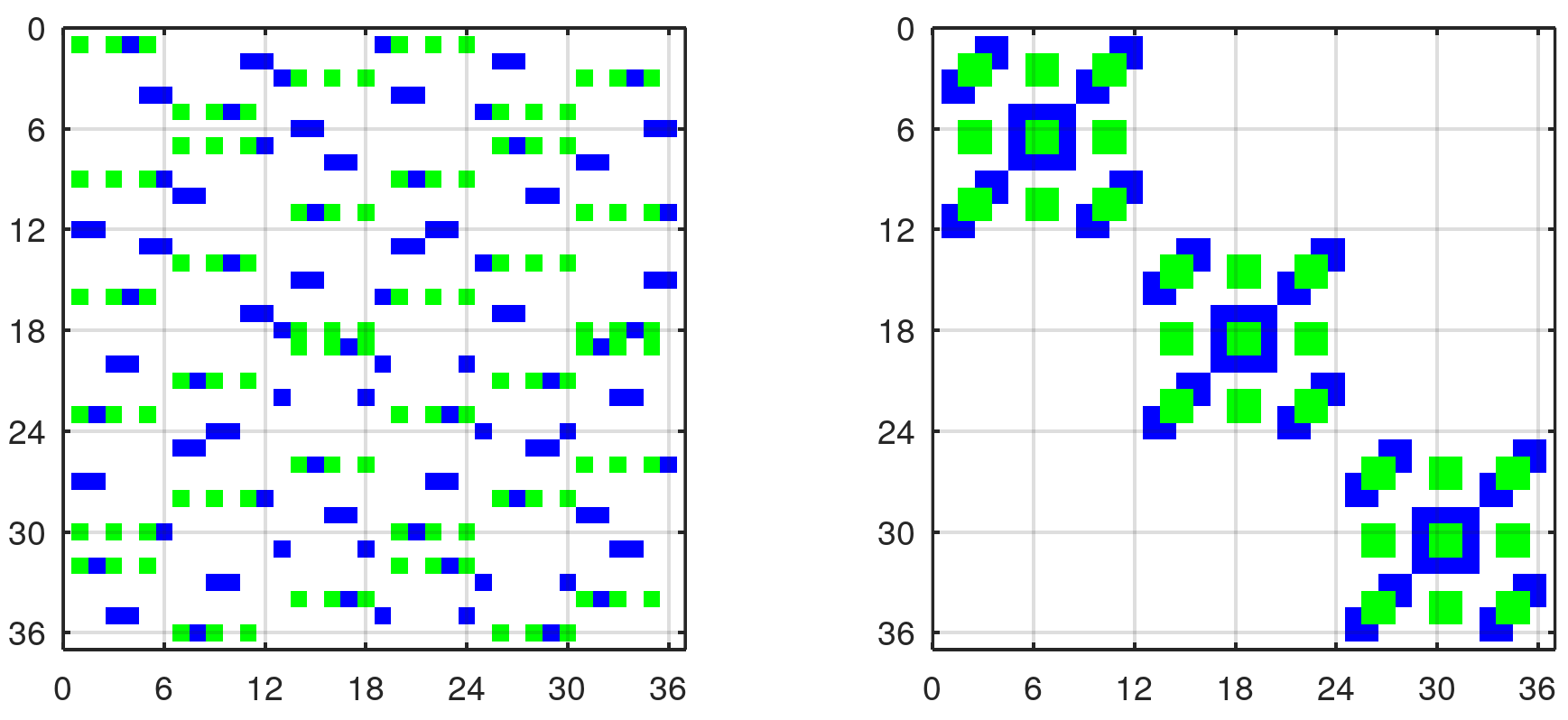}
  \caption{(color online) Left panel: Matrix $Y$ of order $36$,
$2$-unitary after rescaling by $1/6$; result of the map $\mathcal{M}_{\varepsilon}(X)$ for some $X$. Two different colors represent two amplitudes $\in\{\sqrt{3},3\}/6$ of $2\times 108$ non-zero elements. Remaining blank entries denote zeros. Right panel: The same matrix reshuffled and transformed by some two permutation matrices, $P_LY^{\rm\Gamma R}P_R$, in order to form a highly symmetric structure at the price of loss of $2$-unitarity. In both cases, phases are omitted as irrelevant at this stage.}
  \label{fig:Y}
\end{figure}

Another, more important observation is that not every $Y\in\mathbb{Y}$
can be used to minimize the objective function $\mathcal{Z}(Y)$.
Actually, exhaustive numerical search (for many different seeds) revealed only a tiny fraction of $2$-unitary matrices $Y$ for which $\mathcal{Z}\to 0$, assisted by particular $V_j\in\mathbb{U}(6)$. Moreover, in some cases one must additionally realign entries of $Y$ by means of the operations $^{\rm R}$ or $^{\rm\Gamma}$ to make $\mathcal{Z}(Y)$ tend to zero.
One particular example of a matrix $Y$ is shown in Fig.~\ref{fig:Y} (left panel).
This matrix plugged into~\eqref{Z-function} can be transformed into another one, $Y\to_{LU} \mathcal{H}$ for which $|\mathcal{H}_{jk}|=1$ for any value of $j$ and $k$.
Since local unitaries do not affect $2$-unitarity, at this stage we can tentatively and numerically confirm the fact that
there exists a complex Hadamard matrix $\mathcal{H}$ of order $36$, which after rescaling by $1/6$ becomes $2$-unitary.

\section{Structure of the two-unitary CHM}\label{sec:2uCHM}

At first glance the matrix $\mathcal{H}$ does not present any simple structure. However, dephasing (operation that brings any CHM to the {\sl normalized} form in which its first row and first column are all-ones vectors) transforms a number of entries into $6^{\rm th}$-roots of unity.
Unfortunately, dephasing also destroys $2$-unitarity.
But such a form of dephased $\mathcal{H}$ suggests, that this matrix might actually be a Butson-type Hadamard one~\cite{Bu62,Bu63}  shifted by an internal parameterization. 
This assumption is supported by the fact that the defect\footnote{Defect of the matrix $U$, denoted ${\bf d}(U)$, is a natural number (including zero); it indicates the upper bound on the possible number of internal parameters that can be introduced into $U$, without destroying imposed constraints.}~\cite{TZ08} of $\mathcal{H}$ does not vanish; depending on the result of optimization it might take different values, e.g. $61$ (see below), but it never equals to zero.
Indeed, putting some effort,
using unimodularity and orthogonality conditions, 
one can fully recover the analytical form of $\mathcal{H}$, which manifests
itself as a Butson-type CHM with all entries being $6^{\rm th}$-roots of unity, as presumed; $\mathcal{H}\in{\rm B}\mathbb{H}(36, 6)$. An array of integer-valued phases of $\mathcal{H}$ is presented in Appendix~\ref{app:full-form}. Now we can formally arrive at the following
observation, which solves the open problem formulated in Ref.~\cite{BRZ23};

\begin{proposition}
There exists a $2$-unitary complex Hadamard matrix $\mathcal{H}\in\mathbb{H}^{2}(36)$ which is a Butson-type matrix ${\rm B}\mathbb{H}(36,6)$.
\end{proposition}

Non-vanishing defect suggests that $\mathcal{H}$ might not be an isolated point
in the space ${\rm B}\mathbb{H}(36, 6)$ of Butson-type matrices. In fact, we show below that $\mathcal{H}$ admits internal parameterization in the form of $19$ affine parameters plus $5$ non-affine ones. Affinity (vs. non-affinity) indicates that the character of variability of phases in $\mathcal{H}_{jk}$ as functions of orbit parameters, $\alpha_{jk}$, is only linear, i.e.
\begin{equation}
\mathcal{H}_{jk} = \exp\big\{i 2 \pi\varphi_{jk}\big\} \to \mathcal{H}_{jk}(\alpha_{jk}) = \exp\big\{i 2 \pi (\varphi_{jk}+\alpha_{jk})\big\}.\end{equation}
So, in the most general form matrix $\mathcal{H}$ reads $\mathcal{H}=\mathcal{H}(\alpha,\eta)$ with $\alpha\in\mathbb{R}^{19}$ and $\eta\in\mathbb{R}^5$.
However, for the purpose of this paper we shall focus only on affine orbits and properties that can be derived from these additional degrees of freedom, leaving detailed description of the non-affine dependence for a possible future investigation.

\begin{proposition}
Two-unitary CHM matrix $\mathcal{H}$ of order $36$ belongs to a $19$-dimensional affine family.
\end{proposition}
\begin{proof}
Proofs of {\bf Proposition 1} and {\bf 2} are straightforward and reduce to direct examination of the analytic formulas. Full form of the affine family is provided in Appendix~\ref{app:full-form}.
\end{proof}

In the following, let ${\bf 0}$ denote a vector of zeros ${\bf 0}\in\mathbb{R}^{19}$ and
$\mathcal{H} = \mathcal{H}({\bf 0})$.
Hence, we consider $\mathcal{H}(\alpha)$,
with $\alpha\in[0,2\pi)^{\times 19}\setminus\{{\bf 0}\}$.
First of all, any non-zero value of $\alpha$ does not change the properties of being $2$-unitary and CHM, in contrary to local unitary operations which immediately move $\mathcal{H}$ outside the set $\mathbb{H}(36)$.
This makes this result quite important, providing a flexible family of very special AME states, the form of which can be controlled by fine-tuning its degrees of freedom, if one would want to
realize such an object experimentally.

Moreover, special choices of the vector of phases $\alpha$ can simplify the form $\mathcal{H}(\alpha)$ further.
For the sake of simplicity in presentation in the next two examples we assume that $\alpha_j\in\{0,1,2,3,4,5\}$
and actual parameter $\alpha=\exp\{i \pi \alpha_j / 3\}$.
The vector of phases
\begin{equation}
    \sigma=[3, 3, 3, 3, 4, 1, 5, 0, 3, 5, 5, 4, 4, 2, 4, 3, 3, 3, 3]
\end{equation}
turns $\mathcal{H}(\sigma)$ into a symmetric form, $\mathcal{H}(\sigma)=\mathcal{H}(\sigma)^{\rm T}$.
This fact supports the observation from Ref.~\cite{Br23}, where we noticed that vast majority of CHM can be brought to the symmetric or Hermitian form (depending on the dimension). The problem of symmetrizability of CHM extends the same problem for real Hadamard matrices, known in the literature since at least $30$ years and as such it requires independent studies, see Ref.~\cite{MDK15} and references therein. In particular, new examples of symmetric real-valued Hadamard matrices of orders $188$, $292$ and $452$ are constructed in Ref.~\cite{BDK18}.

One can confirm that for any $\varphi\in[0,2\pi)$ and $\gamma=3\varphi/\pi$
the following vector of parameters depending on a single phase
\begin{equation}
    \delta=\delta(\gamma)=[2,5,0,0,0,4,5,3,0,0,0,2,2,\gamma-5,\gamma,\gamma-1,\gamma+5,\gamma-4,\gamma+3],
\end{equation}
provides $\mathcal{H}(\delta)$ with a constant-valued diagonal, i.e. ${\rm arg}(\mathcal{H}(\delta)_{jj})=\varphi$. In particular, suitable values of $\varphi$ allow to obtain matrices
from the class ${\rm B}\mathbb{H}(36,6k)$ with $k\in\mathbb{N}_1$.
Despite many numerical and analytical attempts, no
simpler Butson-type matrix $\mathcal{H}$ was found
and currently six is the smallest possible root of unity that
a $2$-unitary CHM of order $36$ can admit.

By appropriately tuning $\alpha$, it is possible to impose many different
forms of $\mathcal{H}(\alpha)$ depending on the actual requirements. 
Since internal phases cover entire matrix with different intensity, one can adjust a selected subset of entries, optimizing over $19$-dimensional space.

\section{Alternative construction via biunimodular vectors}\label{sec:bu}

In a recent publication~\cite{SAR23}, the authors provide independent construction of a $2$-unitary CHM of order $36$
based on the approach via biunimodular vectors~\cite{FR15}. Such vectors remain unimodular under the transformation
by $F_6\otimes F_6$, where $F_6$ is the standard unitary Fourier matrix of order six.
Three biunimodular vectors $\{\Lambda_j\}_{j=1,2,3}$ of size $36$, listed in Ref.~\cite{SAR23} read:
\begin{align}
\Lambda_1&=\frac{1}{6}[0,1,0,1,3,3,3,3,1,5,2,4,2,1,3,1,2,3,1,1,2,0,3,5,5,3,2,3,2,5,4,4,1,5,5,1],\\
\Lambda_2&=\frac{1}{6}[0,2,3,3,2,0,0,3,2,2,0,4,2,0,3,5,0,0,0,5,0,0,2,0,2,2,5,3,2,4,2,3,0,2,0,0],\\
\Lambda_3&=\frac{1}{3}[0,2,2,0,0,1,0,1,1,1,2,1,0,2,0,2,2,2,2,0,2,2,2,1,1,1,2,0,2,2,0,1,2,2,1,0].
\end{align}
They are building blocks for three matrices $\mathcal{U}_j$ as follows,
\begin{equation}
\mathcal{U}_j=\underbrace{(F_6\otimes\mathbb{I})P(F_6\otimes\mathbb{I})}_{=K}{\rm diag}(\exp\big\{i2\pi\Lambda_j\big\})\underbrace{(F_6\otimes\mathbb{I})P(F_6^{\dagger}\otimes\mathbb{I})}_{=L},
\end{equation}
where the permutation $P$ is a generalized {\rm CNOT} gate (i.e. addition modulo $6$)
\begin{equation}
    P=\sum_{j=0}^{5}|j\rangle\langle j|X^j \quad : \quad X = \sum_{j=0}^5|j\oplus_{\rm mod\, 6} 1\rangle\langle j|
\end{equation}
and matrices $K$ and $L$ are both ${\rm B}\mathbb{H}(36, 6)$ such that 
$K=K^{\dagger}$, $L=L^{\rm T}$ and $K(XA\otimes\mathbb{I})=L,$ where $A$ is 
anti-diagonal permutation matrix. It is straightforward to confirm that indeed all three matrices $\mathcal{U}_j$ are $2$-unitary CHM.\footnote{
It is worth to mention that neither $\mathcal{U}_j$ nor $\mathcal{H}(\alpha)$ were the first instances of 
the $2$-unitary CHM of order $36$. Such a matrix was first obtained numerically by a modified
Sinkhorn algorithm~\cite{BRZ23} with additional unimodularity condition imposed on the matrix elements. 
Soon it was converted to analytical form with additional internal parameterization and put aside as a toy available at: \href{https://chaos.if.uj.edu.pl/~wojtek/A36_6}{\texttt{https://chaos.if.uj.edu.pl/$\sim$wojtek/A36\_6}},
too complicated
for a formal presentation. Eventually, the matrix $\mathcal{H}$ was rediscovered twice in a simplified form in two different ways,
in Ref.~\cite{SAR23} and in this work.}

As all $\mathcal{U}_j$ are characterized by a non-vanishing value of defect, they all might possibly admit
internal parameterizations too. Subsequently, appropriate choice of parameters might cause their orbit to intersect.
A quick instrument to examine whether it is possible to introduce internal affine parameterization is to
check the existence of particularly related columns in a given matrix -- ``ER-pairs''~\cite{Go13} -- a method that necessarily works for objects of even order. This might be accompanied by the method of introducing internal parameterization described in Ref.~\cite{BGZ17}. To this end one must exactly solve a system of linear equations associated with a given matrix, the solution of which indicates the form of an additional matrix $R$ containing affine parameterization.
Using these tools, we were able to obtain conclusive results only for $\mathcal{U}_3$.

\begin{proposition}
There exists a one-parameter affine family
$\mathcal{U}_3(a) = \mathcal{U}_3 \circ \exp\{i 2\pi R\}$, where
$\circ$ is entry-wise product and 
$R=R(a)$ is a matrix of free phases
with only three particular rows affected by parameter $a\in[0,2\pi)$
\begin{align}
R_{2}&=R_{14}=R_{26}=\nonumber\\
&=[\bullet,a,\bullet,\bullet,a,\bullet,\bullet,\bullet,a,\bullet,\bullet,a,a,\bullet,\bullet,a,\bullet,\bullet,\bullet,a,\bullet,\bullet,a,\bullet,\bullet,\bullet,a,\bullet,\bullet,a,a,\bullet,\bullet,a,\bullet,\bullet],
\end{align}
where bullets denote zeros for readability.    
\end{proposition}

Interestingly, the matrix $\mathcal{U}_3$ seems the ``closest'' to the one introduced in this paper due to the value of its defect, ${\bf d}(\mathcal{U}_3)=185$.

Having all these matrices, one question that remains is how much different they are.
In other words, whether it is possible to transform one into another via either local unitary
or global Hadamard-like operations. In the latter case we say that two (complex) Hadamard matrices $H_1$ and $H_2$ are $H$-equivalent\footnote{Here, we do not consider additional transpose-equivalence.}, if one matrix can be unitarily rotated into the other one by means of two monomial unitary matrices $M_1$ and $M_2$, written $H_1\simeq_{\rm H} H_2\Longleftrightarrow H_1=M_1H_2M_2$~\cite{Ha97, LOS20}. Problem of $H$-equivalence is also addressed in Ref.~\cite{Mi14} for a special subset of real Hadamard matrices, called Sylvester matrices. In that case it reduces to permutations and changes the sign of rows and columns of $H$.

\begin{table}[h!]
\begin{center}
\begin{tabular}{ l | r | c | c | c | c}
 matrix $M$ & defect of $M$ & Butson class & symmetries & $2$-unitarity & \# parameters\\ 
 \hline\hline
 $\mathcal{H}({\bf 0})$ & 79  & ${\rm B}\mathbb{H}(36, 6)$ & - & yes & 19\\
 $\mathcal{H}(\alpha)$ & 61 & - & - & yes& 19\\
 $\mathcal{H}(\sigma)$ & 185 & - & $M = M^{\rm T}$ & yes& 19\\
 $\mathcal{H}(\delta)$ & 185 & - & - & yes& 19\\
 \hline
 $\mathcal{U}_1$ & $47$ & ${\rm B}\mathbb{H}(36, 6)$ & unknown & yes& unknown\\
 $\mathcal{U}_2$ & $6$ & ${\rm B}\mathbb{H}(36, 6)$ & unknown & yes& unknown\\
 $\mathcal{U}_3 = \mathcal{U}_3({\bf 0})$ & $185$& ${\rm B}\mathbb{H}(36, 6)$ & unknown & yes& 1\\
 $\mathcal{U}_3(a\neq 0)$ & $\{119, 141, 185\}$ & - & unknown & no & 1\\
\end{tabular}
\caption{\label{tab:matrix-summary}
Characteristics of known Hadamard matrices of order $36$, $2$-unitary up to rescaling. Last column provides lower bound for the number of affine parameters.}
\end{center}
\end{table}
In order to check LU-equivalence one should use the set of invariants
of an analyzed matrix $A$, denoted by
$A(\rho_1,\rho_2,\rho_3,\rho_4)$ described in Ref.~\cite{RRKL23}.
Appropriately chosen permutations $\rho_j\in\mathbb{S}_n$, applied
to rearrange $4$-index in matrix $A\in\mathbb{C}(d^2)$ can be used
to determine the LU-class it belongs to.
In other words, different values of invariants assigned to different matrices imply that they are representatives of different LU-classes.
Numerical calculations suggest that this is the case for $\mathcal{H}({\bf 0})$ and $\mathcal{U}_{1,2,3}$, however numerical complexity implied by the fully non-zero form of the matrices (no vanishing entries), prevents us from any formal statement regarding the general case 
of $\mathcal{H}(\alpha)$ and $\mathcal{U}_3$.

As for the $H$-equivalence we will use
another numerical signature -- the aforementioned defect of a matrix.
Value of the defect, being invariant with respect to monomial transformations, can also
tell whether two Hadamard matrices belong to different (Hadamard) classes, however one must remember that this criterion works in one direction only.
Several attributes of
every matrix have been collected in Table~\ref{tab:matrix-summary}.

Armed only with a numerical evidence we conclude in the following
\begin{conjecture}
    All four matrices $\mathcal{H}(\alpha)$, $\mathcal{U}_{1}$, $\mathcal{U}_{2}$ and $\mathcal{U}_3(a)$ of order $36$ are neither locally unitarily equivalent, nor $H$-equivalent for any values of $(\alpha,a)\in[0,2\pi)^{\times 19}\times [0,2\pi)$.
\end{conjecture}

\section{Summary}\label{sec:summary}

In this paper we constructed a $19$-dimensional affine family of $2$-unitary complex Hadamard matrices of order $36$, which can be further extended over non-affine subspace.
Such a matrix is a new unitary representation of absolutely maximally entangled state of four quhexes.
The advantage of this object over previous solutions is two-fold.
Firstly, it admits only one amplitude while having no zeros.
It is a kind of trade-off with comparison with
the original matrix representing the ``golden'' AME state~\cite{RBBRLZ22}, which contains
rather complicated $20^{\rm th}$-root of unity of three different amplitudes spread over
the matrix with as many as $1184$ vanishing entries.
Secondly, being a member of a multidimensional family preserving both: Hadamardness and the property of being a $2$-unitary matrix, it might serve as a flexible object in practical performance. The problem of preparation an associated quantum circuit with
four pure quhexes at the input and $\mathcal{H}$ or $\mathcal{U}_{1,2,3}$ as output, designed
for a potential experimental realization, is currently subjected to an examination.

\section{Acknowledgements}
We gratefully acknowledge a fruitful cooperation with Suhail Ahmad Rather, especially his numerical recipes that once led to the very first numerical instance of the $2$-unitary CHM of order $36$. It is a pleasure to thank Rafa{\l} Bistro\'{n} and Jakub Czartowski for many inspiring discussions. We are also indebted to anonymous referees for their valuable remarks. WB is supported by NCN through the SONATA BIS grant no. 2019/34/E/ST2/00369. K\.{Z} is supported by NCN under the Quantera project no. 2021/03/Y/ST2/0019.

\appendix

\section{Explicit form of two-unitary CHM}\label{app:full-form}

Full form of the $2$-unitary Hadamard matrix $\mathcal{H}$
depending on the $19$ internal affine parameters is
\begin{equation}
    \mathcal{H}=\mathcal{H}(\alpha)=\exp\Big\{\frac{i\pi}{3} B\Big\}\circ\exp\Big\{\frac{i\pi}{3}A\Big\}\in\mathbb{H}^{\rm 2}(36),
\end{equation}
where $A=A(\alpha)$ is the matrix of order $36$ of internal affine parameters and
both exponential functions acting on matrices should be understood element-wise (as well as the product $\circ$).
The base-matrix $B$ of sixth-roots of unity reads
\begin{equation}
B=\left[
    \begin{array}{c}
1\,\,5\,\,5\,\,5\,\,1\,\,3\,\,4\,\,2\,\,4\,\,4\,\,2\,\,4\,\,4\,\,2\,\,0\,\,0\,\,0\,\,2\,\,1\,\,5\,\,5\,\,5\,\,1\,\,3\,\,0\,\,4\,\,0\,\,0\,\,4\,\,0\,\,0\,\,0\,\,2\,\,4\,\,2\,\,0\\
5\,\,5\,\,5\,\,1\,\,3\,\,1\,\,1\,\,3\,\,3\,\,1\,\,3\,\,3\,\,0\,\,4\,\,4\,\,4\,\,0\,\,2\,\,2\,\,2\,\,2\,\,4\,\,0\,\,4\,\,0\,\,2\,\,2\,\,0\,\,2\,\,2\,\,1\,\,3\,\,5\,\,3\,\,1\,\,1\\
3\,\,3\,\,5\,\,1\,\,5\,\,3\,\,0\,\,0\,\,4\,\,0\,\,0\,\,4\,\,0\,\,0\,\,0\,\,2\,\,4\,\,2\,\,3\,\,3\,\,5\,\,1\,\,5\,\,3\,\,2\,\,2\,\,0\,\,2\,\,2\,\,0\,\,2\,\,4\,\,2\,\,0\,\,0\,\,0\\
1\,\,3\,\,5\,\,3\,\,1\,\,1\,\,3\,\,1\,\,3\,\,3\,\,1\,\,3\,\,2\,\,2\,\,4\,\,0\,\,4\,\,2\,\,4\,\,0\,\,2\,\,0\,\,4\,\,4\,\,2\,\,0\,\,2\,\,2\,\,0\,\,2\,\,3\,\,1\,\,5\,\,5\,\,5\,\,1\\
5\,\,1\,\,5\,\,3\,\,3\,\,3\,\,2\,\,4\,\,4\,\,2\,\,4\,\,4\,\,2\,\,4\,\,0\,\,4\,\,2\,\,2\,\,5\,\,1\,\,5\,\,3\,\,3\,\,3\,\,4\,\,0\,\,0\,\,4\,\,0\,\,0\,\,4\,\,2\,\,2\,\,2\,\,4\,\,0\\
3\,\,1\,\,5\,\,5\,\,5\,\,1\,\,5\,\,5\,\,3\,\,5\,\,5\,\,3\,\,4\,\,0\,\,4\,\,2\,\,2\,\,2\,\,0\,\,4\,\,2\,\,2\,\,2\,\,4\,\,4\,\,4\,\,2\,\,4\,\,4\,\,2\,\,5\,\,5\,\,5\,\,1\,\,3\,\,1\\
4\,\,2\,\,4\,\,1\,\,5\,\,1\,\,5\,\,3\,\,1\,\,4\,\,4\,\,0\,\,3\,\,1\,\,1\,\,4\,\,0\,\,2\,\,0\,\,4\,\,0\,\,3\,\,1\,\,3\,\,1\,\,1\,\,3\,\,2\,\,0\,\,4\,\,3\,\,1\,\,1\,\,4\,\,0\,\,2\\
2\,\,4\,\,4\,\,5\,\,1\,\,1\,\,2\,\,0\,\,0\,\,3\,\,5\,\,1\,\,5\,\,5\,\,5\,\,4\,\,0\,\,4\,\,1\,\,3\,\,3\,\,4\,\,0\,\,0\,\,3\,\,5\,\,1\,\,2\,\,0\,\,0\,\,2\,\,2\,\,2\,\,1\,\,3\,\,1\\
2\,\,2\,\,0\,\,5\,\,5\,\,3\,\,3\,\,3\,\,3\,\,2\,\,4\,\,2\,\,1\,\,1\,\,3\,\,2\,\,0\,\,4\,\,4\,\,4\,\,2\,\,1\,\,1\,\,5\,\,5\,\,1\,\,5\,\,0\,\,0\,\,0\,\,1\,\,1\,\,3\,\,2\,\,0\,\,4\\
0\,\,4\,\,0\,\,3\,\,1\,\,3\,\,0\,\,0\,\,2\,\,1\,\,5\,\,3\,\,3\,\,5\,\,1\,\,2\,\,0\,\,0\,\,5\,\,3\,\,5\,\,2\,\,0\,\,2\,\,1\,\,5\,\,3\,\,0\,\,0\,\,2\,\,0\,\,2\,\,4\,\,5\,\,3\,\,3\\
0\,\,2\,\,2\,\,3\,\,5\,\,5\,\,1\,\,3\,\,5\,\,0\,\,4\,\,4\,\,5\,\,1\,\,5\,\,0\,\,0\,\,0\,\,2\,\,4\,\,4\,\,5\,\,1\,\,1\,\,3\,\,1\,\,1\,\,4\,\,0\,\,2\,\,5\,\,1\,\,5\,\,0\,\,0\,\,0\\
4\,\,4\,\,2\,\,1\,\,1\,\,5\,\,4\,\,0\,\,4\,\,5\,\,5\,\,5\,\,1\,\,5\,\,3\,\,0\,\,0\,\,2\,\,3\,\,3\,\,1\,\,0\,\,0\,\,4\,\,5\,\,5\,\,5\,\,4\,\,0\,\,4\,\,4\,\,2\,\,0\,\,3\,\,3\,\,5\\
3\,\,3\,\,3\,\,5\,\,1\,\,5\,\,2\,\,2\,\,4\,\,0\,\,4\,\,2\,\,0\,\,0\,\,4\,\,0\,\,0\,\,4\,\,5\,\,1\,\,5\,\,3\,\,3\,\,3\,\,2\,\,2\,\,4\,\,0\,\,4\,\,2\,\,4\,\,4\,\,2\,\,4\,\,4\,\,2\\
1\,\,1\,\,3\,\,5\,\,3\,\,1\,\,5\,\,1\,\,3\,\,1\,\,5\,\,5\,\,2\,\,0\,\,2\,\,2\,\,0\,\,2\,\,2\,\,0\,\,4\,\,4\,\,4\,\,0\,\,2\,\,4\,\,0\,\,4\,\,2\,\,2\,\,3\,\,1\,\,3\,\,3\,\,1\,\,3\\
3\,\,5\,\,1\,\,5\,\,3\,\,3\,\,2\,\,4\,\,2\,\,0\,\,0\,\,0\,\,0\,\,2\,\,2\,\,0\,\,2\,\,2\,\,5\,\,3\,\,3\,\,3\,\,5\,\,1\,\,2\,\,4\,\,2\,\,0\,\,0\,\,0\,\,4\,\,0\,\,0\,\,4\,\,0\,\,0\\
1\,\,3\,\,1\,\,5\,\,5\,\,5\,\,5\,\,3\,\,1\,\,1\,\,1\,\,3\,\,2\,\,2\,\,0\,\,2\,\,2\,\,0\,\,2\,\,2\,\,2\,\,4\,\,0\,\,4\,\,2\,\,0\,\,4\,\,4\,\,4\,\,0\,\,3\,\,3\,\,1\,\,3\,\,3\,\,1\\
3\,\,1\,\,5\,\,5\,\,5\,\,1\,\,2\,\,0\,\,0\,\,0\,\,2\,\,4\,\,0\,\,4\,\,0\,\,0\,\,4\,\,0\,\,5\,\,5\,\,1\,\,3\,\,1\,\,5\,\,2\,\,0\,\,0\,\,0\,\,2\,\,4\,\,4\,\,2\,\,4\,\,4\,\,2\,\,4\\
1\,\,5\,\,5\,\,5\,\,1\,\,3\,\,5\,\,5\,\,5\,\,1\,\,3\,\,1\,\,2\,\,4\,\,4\,\,2\,\,4\,\,4\,\,2\,\,4\,\,0\,\,4\,\,2\,\,2\,\,2\,\,2\,\,2\,\,4\,\,0\,\,4\,\,3\,\,5\,\,5\,\,3\,\,5\,\,5\\
1\,\,2\,\,5\,\,2\,\,1\,\,0\,\,0\,\,1\,\,0\,\,3\,\,4\,\,3\,\,0\,\,3\,\,2\,\,1\,\,2\,\,3\,\,1\,\,2\,\,5\,\,2\,\,1\,\,0\,\,4\,\,5\,\,4\,\,1\,\,2\,\,1\,\,4\,\,5\,\,0\,\,3\,\,0\,\,5\\
5\,\,2\,\,5\,\,4\,\,3\,\,4\,\,3\,\,2\,\,5\,\,0\,\,5\,\,2\,\,4\,\,3\,\,2\,\,3\,\,4\,\,1\,\,2\,\,5\,\,2\,\,1\,\,0\,\,1\,\,4\,\,3\,\,0\,\,1\,\,0\,\,3\,\,3\,\,4\,\,1\,\,4\,\,3\,\,2\\
3\,\,0\,\,5\,\,4\,\,5\,\,0\,\,2\,\,5\,\,0\,\,5\,\,2\,\,3\,\,2\,\,1\,\,2\,\,3\,\,0\,\,3\,\,3\,\,0\,\,5\,\,4\,\,5\,\,0\,\,0\,\,3\,\,4\,\,3\,\,0\,\,1\,\,0\,\,3\,\,0\,\,5\,\,4\,\,5\\
1\,\,0\,\,5\,\,0\,\,1\,\,4\,\,5\,\,0\,\,5\,\,2\,\,3\,\,2\,\,0\,\,1\,\,2\,\,5\,\,2\,\,1\,\,4\,\,3\,\,2\,\,3\,\,4\,\,1\,\,0\,\,1\,\,0\,\,3\,\,4\,\,3\,\,5\,\,2\,\,1\,\,0\,\,1\,\,2\\
5\,\,4\,\,5\,\,0\,\,3\,\,0\,\,4\,\,3\,\,0\,\,1\,\,0\,\,3\,\,4\,\,5\,\,2\,\,5\,\,4\,\,3\,\,5\,\,4\,\,5\,\,0\,\,3\,\,0\,\,2\,\,1\,\,4\,\,5\,\,4\,\,1\,\,2\,\,1\,\,0\,\,1\,\,2\,\,5\\
3\,\,4\,\,5\,\,2\,\,5\,\,4\,\,1\,\,4\,\,5\,\,4\,\,1\,\,2\,\,2\,\,5\,\,2\,\,1\,\,0\,\,1\,\,0\,\,1\,\,2\,\,5\,\,2\,\,1\,\,2\,\,5\,\,0\,\,5\,\,2\,\,3\,\,1\,\,0\,\,1\,\,2\,\,5\,\,2\\
0\,\,1\,\,0\,\,0\,\,1\,\,0\,\,1\,\,4\,\,3\,\,5\,\,0\,\,1\,\,3\,\,4\,\,1\,\,1\,\,0\,\,5\,\,4\,\,5\,\,4\,\,4\,\,5\,\,4\,\,5\,\,0\,\,1\,\,1\,\,4\,\,3\,\,3\,\,4\,\,1\,\,1\,\,0\,\,5\\
4\,\,3\,\,0\,\,4\,\,3\,\,0\,\,0\,\,5\,\,4\,\,2\,\,3\,\,0\,\,5\,\,2\,\,5\,\,1\,\,0\,\,1\,\,5\,\,4\,\,1\,\,5\,\,4\,\,1\,\,5\,\,0\,\,3\,\,3\,\,2\,\,1\,\,2\,\,5\,\,2\,\,4\,\,3\,\,4\\
4\,\,1\,\,2\,\,4\,\,1\,\,2\,\,5\,\,4\,\,5\,\,3\,\,0\,\,3\,\,1\,\,4\,\,3\,\,5\,\,0\,\,1\,\,2\,\,5\,\,0\,\,2\,\,5\,\,0\,\,3\,\,0\,\,3\,\,5\,\,4\,\,5\,\,1\,\,4\,\,3\,\,5\,\,0\,\,1\\
2\,\,3\,\,2\,\,2\,\,3\,\,2\,\,4\,\,5\,\,0\,\,0\,\,3\,\,2\,\,3\,\,2\,\,1\,\,5\,\,0\,\,3\,\,3\,\,4\,\,3\,\,3\,\,4\,\,3\,\,3\,\,0\,\,5\,\,1\,\,2\,\,3\,\,0\,\,5\,\,4\,\,2\,\,3\,\,0\\
2\,\,1\,\,4\,\,2\,\,1\,\,4\,\,3\,\,4\,\,1\,\,1\,\,0\,\,5\,\,5\,\,4\,\,5\,\,3\,\,0\,\,3\,\,0\,\,5\,\,2\,\,0\,\,5\,\,2\,\,1\,\,0\,\,5\,\,3\,\,4\,\,1\,\,5\,\,4\,\,5\,\,3\,\,0\,\,3\\
0\,\,3\,\,4\,\,0\,\,3\,\,4\,\,2\,\,5\,\,2\,\,4\,\,3\,\,4\,\,1\,\,2\,\,3\,\,3\,\,0\,\,5\,\,1\,\,4\,\,5\,\,1\,\,4\,\,5\,\,1\,\,0\,\,1\,\,5\,\,2\,\,5\,\,4\,\,5\,\,0\,\,0\,\,3\,\,2\\
5\,\,4\,\,5\,\,0\,\,3\,\,0\,\,2\,\,5\,\,4\,\,3\,\,4\,\,5\,\,4\,\,1\,\,2\,\,1\,\,4\,\,5\,\,3\,\,0\,\,3\,\,2\,\,1\,\,2\,\,2\,\,5\,\,4\,\,3\,\,4\,\,5\,\,0\,\,3\,\,4\,\,3\,\,0\,\,1\\
5\,\,0\,\,1\,\,4\,\,1\,\,0\,\,5\,\,4\,\,3\,\,4\,\,5\,\,2\,\,0\,\,1\,\,0\,\,3\,\,4\,\,3\,\,4\,\,1\,\,0\,\,5\,\,0\,\,1\,\,2\,\,1\,\,0\,\,1\,\,2\,\,5\,\,5\,\,0\,\,5\,\,2\,\,3\,\,2\\
5\,\,0\,\,3\,\,0\,\,5\,\,4\,\,2\,\,1\,\,2\,\,3\,\,0\,\,3\,\,4\,\,3\,\,0\,\,1\,\,0\,\,3\,\,3\,\,2\,\,1\,\,2\,\,3\,\,0\,\,2\,\,1\,\,2\,\,3\,\,0\,\,3\,\,0\,\,5\,\,2\,\,3\,\,2\,\,5\\
5\,\,2\,\,5\,\,4\,\,3\,\,4\,\,5\,\,0\,\,1\,\,4\,\,1\,\,0\,\,0\,\,3\,\,4\,\,3\,\,0\,\,1\,\,4\,\,3\,\,4\,\,5\,\,2\,\,5\,\,2\,\,3\,\,4\,\,1\,\,4\,\,3\,\,5\,\,2\,\,3\,\,2\,\,5\,\,0\\
5\,\,2\,\,1\,\,0\,\,1\,\,2\,\,2\,\,3\,\,0\,\,3\,\,2\,\,1\,\,4\,\,5\,\,4\,\,1\,\,2\,\,1\,\,3\,\,4\,\,5\,\,2\,\,5\,\,4\,\,2\,\,3\,\,0\,\,3\,\,2\,\,1\,\,0\,\,1\,\,0\,\,3\,\,4\,\,3\\
5\,\,4\,\,3\,\,4\,\,5\,\,2\,\,5\,\,2\,\,5\,\,4\,\,3\,\,4\,\,0\,\,5\,\,2\,\,3\,\,2\,\,5\,\,4\,\,5\,\,2\,\,5\,\,4\,\,3\,\,2\,\,5\,\,2\,\,1\,\,0\,\,1\,\,5\,\,4\,\,1\,\,2\,\,1\,\,4\\
    \end{array}\right].
\end{equation}
Matrix of parameters $A=A(\alpha)$ requires step-by-step presentation
because of technically cumbersome although conceptually simple structure.
Let
\begin{equation}
\alpha=[a,b,c,d,e,f,g,h,i,j,k,l,m,n,o,p,q,r,s]\in[0,2\pi)^{\times 19}.    
\end{equation}
First, we notice that matrix $A$ consists of four identical sub-blocks $C$ of size $18\times 18$,
such that $C=C_{j,k}-C_{l,m}$ with
\begin{equation}
C_{j,k}=\left[
    \begin{array}{cccccccccccccccccc}
j&j&.&j&j&.&j&j&.&j&j&.&j&j&.&j&j&.\\
k&k&.&k&k&.&k&k&.&k&k&.&k&k&.&k&k&.\\
j&j&.&j&j&.&j&j&.&j&j&.&j&j&.&j&j&.\\
k&k&.&k&k&.&k&k&.&k&k&.&k&k&.&k&k&.\\
j&j&.&j&j&.&j&j&.&j&j&.&j&j&.&j&j&.\\
k&k&.&k&k&.&k&k&.&k&k&.&k&k&.&k&k&.\\
j&.&j&j&.&j&j&.&j&j&.&j&j&.&j&j&.&j\\
k&.&k&k&.&k&k&.&k&k&.&k&k&.&k&k&.&k\\
j&.&j&j&.&j&j&.&j&j&.&j&j&.&j&j&.&j\\
k&.&k&k&.&k&k&.&k&k&.&k&k&.&k&k&.&k\\
j&.&j&j&.&j&j&.&j&j&.&j&j&.&j&j&.&j\\
k&.&k&k&.&k&k&.&k&k&.&k&k&.&k&k&.&k\\
.&j&j&.&j&j&.&j&j&.&j&j&.&j&j&.&j&j\\
.&k&k&.&k&k&.&k&k&.&k&k&.&k&k&.&k&k\\
.&j&j&.&j&j&.&j&j&.&j&j&.&j&j&.&j&j\\
.&k&k&.&k&k&.&k&k&.&k&k&.&k&k&.&k&k\\
.&j&j&.&j&j&.&j&j&.&j&j&.&j&j&.&j&j\\
.&k&k&.&k&k&.&k&k&.&k&k&.&k&k&.&k&k\\
    \end{array}\right]
\end{equation}
and
\begin{equation}
C_{l,m}=\left[
    \begin{array}{cccccccccccccccccc}
l&.&.&l&.&.&l&.&.&l&.&.&l&.&.&l&.&.\\
m&.&.&m&.&.&m&.&.&m&.&.&m&.&.&m&.&.\\
l&.&.&l&.&.&l&.&.&l&.&.&l&.&.&l&.&.\\
m&.&.&m&.&.&m&.&.&m&.&.&m&.&.&m&.&.\\
l&.&.&l&.&.&l&.&.&l&.&.&l&.&.&l&.&.\\
m&.&.&m&.&.&m&.&.&m&.&.&m&.&.&m&.&.\\
.&.&l&.&.&l&.&.&l&.&.&l&.&.&l&.&.&l\\
.&.&m&.&.&m&.&.&m&.&.&m&.&.&m&.&.&m\\
.&.&l&.&.&l&.&.&l&.&.&l&.&.&l&.&.&l\\
.&.&m&.&.&m&.&.&m&.&.&m&.&.&m&.&.&m\\
.&.&l&.&.&l&.&.&l&.&.&l&.&.&l&.&.&l\\
.&.&m&.&.&m&.&.&m&.&.&m&.&.&m&.&.&m\\
.&l&.&.&l&.&.&l&.&.&l&.&.&l&.&.&l&.\\
.&m&.&.&m&.&.&m&.&.&m&.&.&m&.&.&m&.\\
.&l&.&.&l&.&.&l&.&.&l&.&.&l&.&.&l&.\\
.&m&.&.&m&.&.&m&.&.&m&.&.&m&.&.&m&.\\
.&l&.&.&l&.&.&l&.&.&l&.&.&l&.&.&l&.\\
.&m&.&.&m&.&.&m&.&.&m&.&.&m&.&.&m&.\\
    \end{array}\right]
\end{equation}
and then
\begin{equation}
A=\left[\begin{array}{cc}
C & C\\
C & C
\end{array}
\right].
\end{equation}
Next, such $A$ is globally affected at the following positions:
\begin{align}
A_{jk} = a \quad &:\quad (j,k)\in[1,36]\times ([1,6]\cup[19,24]),\\
A_{jk} = b \quad &:\quad (j,k)\in[1,36]\times ([13,18]\cup[31,36]),\\
A_{jk} = c \quad &:\quad (j,k)\in[1,36]\times \{1, 7, 13, 19, 25, 31\},\\
A_{jk} = d \quad &:\quad (j,k)\in[1,36]\times \{2, 8, 14, 20, 26, 32\},\\
A_{jk} = e \quad &:\quad (j,k)\in[1,36]\times \{3, 9, 15, 21, 27, 33\},\\
A_{jk} = f \quad &:\quad (j,k)\in[1,6]\times \bigcup_{t\equiv 1\bmod 3}\{t\}\cup[19,24]\times \bigcup_{t\equiv 1\bmod 3}\{t\},\label{union_f}\\
A_{jk} = g \quad &:\quad (j,k)\in[1,6]\times \bigcup_{t\equiv 2\bmod 3}\{t\}\cup[19,24]\times \bigcup_{t\equiv 2\bmod 3}\{t\},\label{union_g}\\
A_{jk} = h \quad &:\quad (j,k)\in[13,18]\times \bigcup_{t\equiv 2\bmod 3}\{t\}\cup[31,36]\times \bigcup_{t\equiv 2\bmod 3}\{t\},\label{union_h}\\
A_{jk} = i \quad &:\quad (j,k)\in[13,18]\times \bigcup_{t\equiv 1\bmod 3}\{t\}\cup[31,36]\times \bigcup_{t\equiv 1\bmod 3}\{t\},\label{union_j}\\
A_{jk} = n \quad &:\quad (j,k)\in\{1, 7, 13, 20, 26, 31\}\times [1,36],\\
A_{jk} = o \quad &:\quad (j,k)\in\{2, 8, 14, 21, 27, 32\}\times [1,36],\\
A_{jk} = p \quad &:\quad (j,k)\in\{3, 9, 15, 22, 28, 33\}\times [1,36],\\
A_{jk} = q \quad &:\quad (j,k)\in\{4, 10, 16, 23, 29, 34\}\times [1,36],\\
A_{jk} = r \quad &:\quad (j,k)\in\{5, 11, 17, 24, 30, 35\}\times [1,36],\\
A_{jk} = s \quad &:\quad (j,k)\in\{6, 12, 18, 25, 31, 36\}\times [1,36].
\end{align}
Eventually, matrix $A$ takes the final form depending on $19$ phases $A=A(a,b,c,...,s)=A(\alpha)$.

\medskip 

\noindent Scripts in \texttt{MATLAB} and \texttt{Mathematica} can be found on \texttt{GitHub} repository~\cite{GitHub}.

\end{document}